\documentclass[aoas]{imsart}

\usepackage{amsmath,amsfonts, graphicx, algorithm,algpseudocode, natbib, comment}

\newcommand\cites[1]{\citeauthor{#1}'s\ (\citeyear{#1})}

\newcommand{\E}{\mathbb{E}}

\begin{document}
\begin{frontmatter}
\title{A Variational EM Method for Mixed Membership Models with Multivariate Rank Data: an Analysis of Public Policy Preferences}
\runtitle{Mixed Membership Models for Rank Data}
\begin{aug}
  \author{\fnms{Y. Samuel} \snm{Wang}\ead[label=e1]{ysamwang@uw.edu}},
  \author{\fnms{Ross L.} \snm{Matsueda}\ead[label=e2]{matsueda@uw.edu}}
  \and
  \author{\fnms{Elena A.}  \snm{Erosheva} \ead[label=e3]{erosheva@uw.edu}}
  
  \runauthor{YS Wang, Matsueda and Erosheva}

  \affiliation{University of Washington}
  \address{University of Washington\\
  Dept of Statistics\\
  Seattle, WA 98195 USA \\ 
            \printead{e1,e3}}
\address{University of Washington\\
  Dept of Sociology\\
  Seattle, WA 98195 USA \\ 
	\printead{e2}}
            
\end{aug}

\begin{abstract}
In this article, we consider modeling ranked responses from a heterogeneous population. Specifically, we analyze data from the Eurobarometer 34.1 survey regarding public policy preferences towards drugs, alcohol and AIDS. Such policy preferences are likely to exhibit substantial differences within as well as across European nations reflecting a wide variety of cultures, political affiliations, ideological perspectives and common practices. We use a mixed membership model to account for multiple subgroups with differing preferences and to allow each individual to possess partial membership in more than one subgroup. Previous methods for fitting mixed membership models to rank data in a univariate setting have utilized an MCMC approach and do not estimate the relative frequency of each subgroup. We propose a variational EM approach for fitting mixed membership models with multivariate rank data. Our method allows for fast approximate inference and explicitly estimates the subgroup sizes. Analyzing the Eurobarometer 34.1 data, we find interpretable subgroups which generally agree with the ``left vs right" classification of political ideologies.  
\end{abstract}

\begin{keyword}
\kwd{mixed membership}
\kwd{rank data}
\kwd{variational inference}
\kwd{eurobarometer}
\kwd{public policy}
\end{keyword}

\end{frontmatter}

\section{Introduction}
Rank data often arise from a heterogeneous population with individuals whose preferences may vary widely. In this article, we consider one such example, public health policy data from the Eurobarometer 34.1. We develop a computationally efficient variational EM procedure to estimate mixed membership models with rank data. In addition to the computational aspects, we also extend the current literature by explicitly estimating the subgroup relative frequencies and accommodating multivariate ranked data.

\section{Public Policy Preferences}
Social scientists have long held that in democratic societies, public opinion plays an important role in the formation of public policies about important social problems \citep[e.g.][]{burstein1998bringing,brooks2008welfare}. Public opinion polls, which provide a window on the attitudes and perspectives of a nation's citizenry, at times elicit ranked data about specific public policies. An example is the Eurobarometer 34.1, a survey commissioned in 1990 to study European perspectives on various political and public health issues \citep{Reif2001}. The Eurobarometer 34.1 collected data on a broad range of topics, using a range of question formats, including measures of health behavior, knowledge of illegal drugs, descriptions of family structure, and attitudes toward children. In particular, three of the survey questions, shown in Figure \ref{fig:questions}, asked about public policy priorities toward addressing societal/public health problems. Therefore, our analysis focuses on examining responses to these questions. We should note that, if the survey contained other pertinent questions of binary or multinomial responses, these data could have been included in the analysis using the \texttt{mixedMem} R package \citep{wang2015mixedMem, rLang} which allows for multivariate analysis when the variables are of different distributions. 

The specific variables we consider address illicit drugs, alcoholism, and AIDS. The survey respondents were asked to rank, in order of priority, policies such as punishment for offenders, information campaigns to educate the public, rehabilitation and treatment, funding of research into causes and treatment, and fighting social causes. Such priority rankings are likely to vary across respondents within a nation, reflecting dissensus among a citizenry, as well as across nations, reflecting national differences in culture, political affiliation, ideological perspective, and common practices.

\begin{figure}[h] 
	\centering \def\arraystretch{1.15}
	\begin{tabular}{|p{13cm}|}
	\hline \tiny 
	\textbf{Question 28: }
	There are various actions that could be taken to eliminate the drugs problem. In your opinion, what is the first priority? And the next most urgent? (Ask respondent to rank all 7, with 1 as the most urgent)
	\begin{enumerate} \tiny \setlength\itemsep{-.25em}
	\item Information campaigns about the dangers of drugs
	\item Hunting down drug pushers and distributors 
	\item Legal penalty for drug taking 
	\item Looking after and treating drug addicts and rehabilitating them 
	\item Funding research into drug substitutes, and into the treatment of drug addiction 
	\item Fighting the social causes of drug addiction
	\item Reinforcing the control or distribution and usage of addictive medicines
	\end{enumerate} 
	\\ \hline \tiny
	\textbf{Question 39: }
	There are various actions that could be taken in order to ease the problem of alcoholism and its consequences. Looking at this card, which is the main priority in your view? and the next? (Rank up to 5)
		\begin{enumerate} \tiny \setlength\itemsep{-.25em}
		\item Information campaigns about the danger of alcoholism 
		\item Stiffer penalties for offenses committed under the influence of alcohol 
		\item Banning advertising for alcoholic drinks 
		\item Increasing taxes on alcohol
		\item Restricting the sale of alcohol. Especially to young people 
		\item Putting lower legal limits on alcohol content
		\item Making social outcasts of alcoholics 
		\item Helping alcoholics to submit ``drying out" 
		\item Funding medical research to develop more effective treatments
		\item Setting up more reception centers, drying out treatment centers
		\end{enumerate} 
		\\ \hline \tiny
		\textbf{Question 47: }	
	There are various actions that could be taken in order to eliminate the problem of AIDS or at least to slow down its development. Looking at this card, which is the main priority in your view? And the next? (Rank all by giving a number from 1 to 5. with 1 as a top priority)
		\begin{enumerate} \tiny \setlength\itemsep{-.25em}
		\item Information campaign about the danger
		\item Punishment for behavior which increases the risk
		\item Identifying and isolating those with AIDS or those who are HIV positive
		\item Treating of those with AIDS and looking after them
		\item Funding research to find a vaccine
		\end{enumerate} \\\hline 
	\end{tabular}
\caption{The three questions of interest from Eurobarometer 34.1 regarding illegal drugs, alcoholism, and AIDS. The survey was administered in Belgium, Denmark, East Germany, France, Greece, Ireland, Italy, Luxembourg, The Netherlands, Portugal, Spain, UK, and West Germany. Further demographic information is given in Table \ref{tab:Demo}.\label{fig:questions}}
\end{figure}

\begin{table}[htbp!]
\centering
\tiny
\caption{\label{tab:topFive}The top 5 observed rankings for each question. The place in the permutation represents the ranking level and the recorded number indicates the order in which the policy option was presented in the questionnaire. See Figure \ref{fig:questions}, for the corresponding policies.}

\small
\begin{tabular}{cc}
  \hline 
Drug & Count \\ 
  \hline
1,2,3,4,5,6,7 & 109 \\ 
  1,2,4,5,6,7,3 &  82 \\ 
  2,1,4,5,6,7,3 &  69 \\ 
  2,1,4,6,7,5,3 &  55 \\ 
  2,3,1,4,5,6,7 &  53 \\ 

   \hline
\end{tabular}
\begin{tabular}{cc}
  \hline
Alcohol& Count\\ 
  \hline
1,2,8,9,10 &  85 \\ 
  1,5,8,9,10 &  61 \\ 
  1,8,9,10,3 &  60 \\ 
  1,2,5,8,10 &  57 \\ 
  1,8,9,10,2 &  53 \\ 
   \hline
\end{tabular}
\begin{tabular}{cc} 
  \hline
AIDS & Count \\ 
  \hline
1,5,4,2,3 & 740 \\ 
  5,4,1,2,3 & 722 \\ 
  5,1,4,2,3 & 670 \\ 
  1,5,4,3,2 & 604 \\ 
  1,4,5,2,3 & 515 \\ 
   \hline
\end{tabular}
\end{table}


The top 5 observed responses for each question, shown in Table \ref{tab:topFive}, are suggestive of significant population heterogeneity. For example, for illegal drugs, a ``legal penalty for drug taking" (policy 3) is highly ranked in the first and fifth most observed permutations, but ranked last in the second, third and fourth most observed permutations. This heterogeneity is not surprising because the individuals in the survey come from a wide variety of nationalities, religious backgrounds and age groups (shown in Table \ref{tab:Demo}). In analyzing such heterogeneity, an important question is whether there are underlying subgroups or policy profiles among citizens. For example, when asked how they prioritize policies about the problem of illegal drugs, do citizens form a single group that varies along a dimension of punishment to rehabilitation, which reflects a conservative-liberal continuum? Or do responses reflect subgroups of citizens, in which some favor punitive measures, others rehabilitation, and still others education? Do some subgroups systematically oppose some policy measures, while favoring others? Moreover, given such subgroups, are some citizens members of multiple groups, favoring, for example, both rehabilitation and information? In modern democracies characterized by a diverse citizenry, such subgroups may be likely to exist. Failure to consider such subgroups, when they in fact exist, may lead to a distorted view of a nation's public. Appropriately modeling the population heterogeneity may be particularly important when the questions cover a wide range of topics, as they do here. 


\begin{table}
\tiny
\caption{\label{tab:Demo} Demographic breakdown of Eurobarometer 34.1 participants}
\begin{tabular}{lr||lr||lr}
  \hline
Nation & Count & Nation & Count & Nation & Count\\ 
  \hline
Belgium & 812 & Greece & 980 & Northern Ireland & 267\\ 
  Denmark & 950 & Ireland & 879 & Portugal & 978  \\ 
  East Germany & 938 & Italy & 1048 & Spain & 864  \\ 
  France & 967 & Luxembourg & 242 &   West Germany & 975  \\ 
  Great Britain & 956 & Netherlands & 1016 & & \\ 
   \hline\\
\end{tabular}
\begin{tabular}{lr||lr}
  \hline
 Religion & Count & Religion & Count \\ 
  \hline
Buddhist &  11 & Orthodox & 1057 \\ 
  Hindu &   2 &   Other & 162 \\ 
  Jewish &  20 & Protestant & 2182\\ 
  Muslim &  23 & Roman Catholic & 5617 \\ 
  None & 2692 & & \\ 
   \hline
\end{tabular} \quad
\begin{tabular}{rr}
  \hline
Age Group & Count \\ 
  \hline
15-24 Years & 2401 \\ 
  25-39 Years &3489 \\ 
  40-54 Years & 2780 \\ 
  55+ Years & 3202 \\ 
  &\\
   \hline
\end{tabular}
\end{table}

\section{Mixed Membership Models}
\subsection{Previous Work}
Many approaches for modeling rank data have been proposed; for a review, see \citet{marden1996analyzing}. In this paper, we focus on the Plackett-Luce model due to several attractive attributes (discussed further in Section \ref{sec:rankData}). Assuming population homogeneity, \citet{hunter2004mm} develop a minorization-maximization method for estimating MLEs for a single Plackett-Luce distribution, and \citet{guiver2009bayesian} present a Bayesian framework for estimation.

Most of the previous work that address heterogeneity in rank data consider the univariate case and specify a mixture model. \cite{gormley2006analysis} assume a mixture of Plackett-Luce distributions, while \cite{busse2007cluster} assume a mixture of Mallow's distributions. In addition, Bayesian non-parametric approaches have been used to allow for an infinite number of latent subgroups \citep{meila2012dirichlet} and an infinite set of alternatives \citep{caron2014bayesian}. \cite{gormley2008mixture} propose a mixture of experts model where individual level covariates specify the probability that an individual belongs to a specific subgroup. However, each of these mixture model approaches assume that every individual always expresses preferences consistent with only a single subgroup. In many cases this may be overly restrictive.

\citet{gormley2009grade} propose a mixed membership model for univariate rank data which allows for intra-subgroup mixing between ranking levels. 
Mixed membership models extend mixture models by allowing an individual membership to be split among multiple subgroups \citep[e.g.][]{airoldi2014handbook}. As \citet[pg 128]{grossManriqueVallier} point out, the structure of a mixed membership model is consistent with \cites{zaller1992nature} model of responses to public opinion polls, in which ``respondents randomly sample from a number of privately held `considerations' relevant to the question at hand.'' \cite{grossManriqueVallier} analyzed data on political ideology, examining core beliefs and values, finding that mixed membership models reveal hidden structures in political ideologies that previous factor analytic results missed.

Direct maximum likelihood estimation of mixed membership models is generally intractable, so MCMC or approximate inference techniques are used. In particular, \citet{gormley2009grade} develop a Metropolis-within-Gibbs sampler. Although the MCMC method allows for direct sampling from the posterior, it scales poorly. Because of the individual level parameters, the total number of parameters is directly proportional to sample size and the number of variables, which can result in slow mixing even for moderate sample sizes. MCMC methods can also require a considerable amount of human effort since convergence diagnostics must be checked for all parameters \citep{gill2008partial}.

\subsection{Contribution of this work}
In this article, we propose a variational EM approach which scales well with the number of observed individuals and is capable of tractably estimating mixed membership models for rank data with a much larger sample size than can be handled by existing MCMC methods. Variational inference has been used as an alternative estimation procedure in mixed membership models where MCMC methods would be computationally infeasible \citep{bleiLDA, erosheva2004mixed, airoldi2009mixed} and has been shown empirically to provide results similar to MCMC in some cases \citep{erosheva2007describing}. A direct computational comparison between the proposed variational method and the MCMC method detailed by \citet{gormley2009grade} is provided in the supplement \citep{wang2015variational:supp}. 

Motivated by Eurobarometer data on public policy priorities, we extend the method of \citet{gormley2009grade} to allow for multivariate data and directly estimate the relative frequencies of each subgroup. The direct estimation of the subgroup relative frequencies can be viewed as an empirical Bayes type procedure or robust modeling practice and has been shown to improve predictive performance in many cases \citep{wang2015general}. Indeed, we show in the supplement that direct estimation of the subgroup relative frequencies drastically improves the goodness-of-fit for exit poll data gathered during the 1997 Irish Presidential Election \citep{wang2015variational:supp}. Finally, informed by exploratory data analyses, we extend the model specification to include two pre-specified subgroups.

The remainder of the article is structured as follows. We introduce the Plackett-Luce distribution in Section \ref{sec:rankData} and the mixed membership rank data model in Section \ref{sec:genModel}. In Section \ref{sec:method}, we review the variational approximation framework and detail the variational EM algorithm. Finally, we use the proposed method to analyze public policy priorities from the Eurobarometer 34.1 survey in Section \ref{sec:interpretation} and conclude with discussion in Section \ref{sec:discussion}.

\section{Modeling Rank Data} \label{sec:rankData}
First, we consider univariate rank data. Suppose there are $V$ alternatives in the choice set (items to be ranked). For each individual $i$, a single observation $X_i$ is a permutation of $N_{i} \leq V$ of the alternatives in the choice set. Each alternative $v\in V$ (in this context, alternatives correspond to policies) is assigned a non-negative support parameter $\theta_v$ which governs how strongly it is preferred to other alternatives \citep{plackett1975analysis}. The support parameters sum to 1 for identifiability. At each ranking level, one of the remaining alternatives is selected with probability proportional to its support parameter. The mass function for the Plackett-Luce model is defined as
\begin{equation}\label{eq:plDistro} \footnotesize
P(X_i) = \prod_{n=1}^{N_i}\frac{\theta_{a(n)_i}}{1-\sum_{c =0}^{n-1}\theta_{a(c)_i}},
\end{equation}
where $a(n)_i$ indicates the alternative selected at the $n^{th}$ ranking level, and $\theta_{a(0)} =0$. The Plackett-Luce selection process can be thought of as a multinomial without replacement, and the support parameters represent the probability of a policy being selected as the top priority.

Although there are various distributions for modeling rank data, the Plackett-Luce model has several attractive properties. First, it can accommodate incomplete rankings when not all alternatives in the choice set are selected. Assuming unranked policies are less preferred than ranked policies, the mass function in Equation \ref{eq:plDistro} has marginalized out any unranked policies. Second, the model satisfies Luce's Choice Axiom \citep{luce1977choice} which states that an individual's relative preference between alternatives $v$ and $v'$ should not change when a third alternative $v''$ is introduced \citep{sen1970collective}. Finally, the parameter space of the Plackett-Luce model is a continuous set in the $V-1$ simplex, while the parameter space of other popular models (notably Mallow's model) may be discrete sets which can greatly complicate inference. 

\section{Generative Model: Multivariate Rank Data} \label{sec:genModel}
\begin{figure}[H]
\centering
\includegraphics[scale = .25]{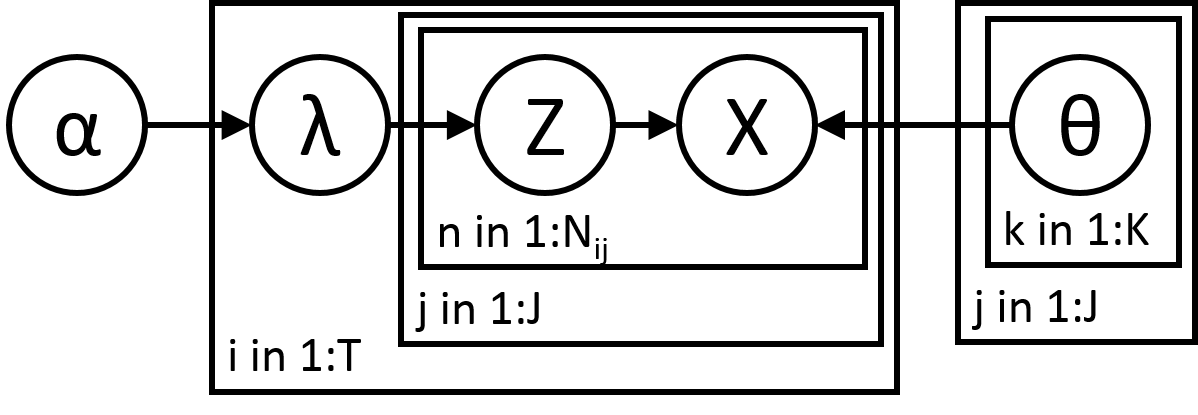} 
\caption{Plate notation of the Multivariate Mixed Membership Rank Data Model\label{fig:plate}. Quantities $\lambda$, Z and X are specific to individuals while quantities $\alpha$ and $\theta$ are global.}
\end{figure}

We propose the following generative model for multivariate rank data. Assuming $K$ latent subgroups, Dirichlet membership parameter $\alpha \in \mathbb{R}^K_{>0}$, $J$ variables each with $V_j$ alternatives, and a set of support parameters $\theta_{jk}$ for each variable $j$ and subgroup $k$ where $\sum_v\theta_{jkv} = 1$, the generative mixed membership model is:
\begin{enumerate} \footnotesize
\item For each individual $i = 1,2\ldots T$ 
\begin{itemize}
\item[1.] Draw a membership vector $\lambda_i \sim$ Dirichlet($\alpha$).
\item[2.] For each variable $j = 1,2\ldots J$
	and for each ranking level $ n = 1,2,\ldots N_{ij}$ 
		\begin{enumerate}
		\item[(a)] Draw a context indicator $Z_{ijn} \sim$ multinomial(1,$\lambda_i$).
		\item[(b)] Draw a policy priority $X_{ijn} \sim$ $\text{Plackett-Luce}(\theta, Z_{ijn} | \{X_{ijc}\}_{c < n})$. 
		\end{enumerate}
	\end{itemize} 
\end{enumerate}
 where $\text{Plackett-Luce}(\theta, Z_{ijn}| \{X_{ijc}\}_{c < n})$ denotes that $X_{ijn}$ is dependent on $X_{ijc}$ for $c < n$ since the alternatives selected at a prior ranking levels cannot be selected again. The corresponding complete data likelihood and the graphical representation are shown in Equation \ref{eq:fullLike} and in Figure \ref{fig:plate}.
\begin{equation} \footnotesize
\begin{aligned}\label{eq:fullLike}
P(X, \Lambda, Z|\alpha, \theta) =\prod^T_i \left\{ \text{Dir}(\lambda_i|\alpha)\prod_j^J \prod_{n}^{N_{ij}}\left( \text{mult}(Z_{ijn}|\lambda_i) \prod_k^K \left[\frac{\theta_{jk a(n)_{ij}}}{1-\sum_{c =0}^{n-1}\theta_{jka(c)_{ij}}}\right]^{Z_{ijnk}} \right)\right\}
\end{aligned}  \end{equation}

In the model, $\lambda_{ik}$ denotes the degree of membership of individual $i$ within subgroup $k$, and in the Eurobarometer context, $\lambda_{ik}$ indicates an individual's level of adherence to a policy ideology. $Z_{ijn}$ is the subgroup governing individual's $i$ selection for variable $j$ at ranking level $n$. Note that the model assumes mixing of subgroup preferences occur both between different variables and within a single observed ranking. Thus, an individual may select their top choice according to the preferences of one subgroup, but select their second choice according to the preferences of another subgroup. We note that, conditional on the membership $\lambda_i$, there is no further dependence enforced on each $Z_{ijn}$ across the ranking levels of a single variable. Although the $Z_{ijn}$ are exchangeable in the generative model, in the posterior conditioned on the observations, the context indicators are no longer exchangeable because of the assumed sequential nature of the ranking procedure.

\section{Variational EM Approach} \label{sec:method}
\subsection{Variational Approximation}
Calculating the marginal probability $P(X|\alpha, \theta) = \int_{Z,\Lambda} P(X, \Lambda, Z | \theta, \alpha)$ requires marginalizing over the simplicial membership vectors $\Lambda$ and context indicators $Z$. Because this calculation is intractable, we use a mean-field variational method which approximates the true posterior for latent variables $\Lambda$ and $Z$. A detailed tutorial of variational inference is provided by \cite{wainwright2008graphical}. 

The variational distribution is 
\begin{equation}\label{eq:varDistro} \footnotesize
Q(Z,\Lambda|\phi, \delta) = \prod_i^T \left( \text{Dir}\left(\lambda_i|\phi_i\right)\prod_j^J \prod_n^{N_{ij}}\text{mult}\left(Z_{ijn}|\delta_{ijn}\right) \right) \end{equation}
with variational parameters $\phi$ and $\delta$ ($\phi_i \in \mathbb{R}^K_{>0}$ and $\delta_{ijn}$ lies in the $K-1$ dimension simplex). Because it factors easily into functions of the variational parameters, this approximation facilitates tractable computation via the Variational EM algorithm. We first derive a lower bound on the marginal distribution of the observed rankings using Jensen's inequality. 
\begin{equation} \small
 \label{eq:Jensen2}
\begin{aligned}
\log \left[p(X|\alpha, \theta) \right] &= \log \left[\int_{\Lambda, Z} \frac{Q(Z, \Lambda|\phi, \delta)}{Q(Z,\Lambda|\phi, \delta)} P(X,Z,\Lambda|\alpha, \theta) \right] \\ 
&= \log\E_Q\left[\frac{P(X,Z,\Lambda |\alpha, \theta)}{Q(Z,\Lambda|\phi, \delta)}\right]\\
 &\geq \E_Q\left\{\log\left[P(X,Z, \Lambda|\alpha, \theta)\right]\right\} - \E_Q\left\{\log\left[Q(Z, \Lambda|\phi, \delta)\right]\right\}
\end{aligned}
\end{equation}
The last line in Equation \ref{eq:Jensen2} is often called the \textbf{E}vidence \textbf{L}ower \textbf{Bo}und (ELBO) and is a function of the data, the variational parameters, $\phi$ and $\delta$, as well as the global parameters $\alpha$ and $\theta$. It can be shown that maximizing the ELBO with respect to the variational parameters $\phi$ and $\delta$ minimizes the KL-divergence between the true posterior and the variational distribution. In addition, fixing the variational parameters and maximizing the lower bound with respect to $\alpha$ and $\theta$ can be used as a surrogate procedure for selecting maximum likelihood estimates for $\alpha$ and $\theta$ \citep{beal2003variational}. Ultimately, by maximizing the lower bound (the ELBO), we simultaneously find pseudo-MLE estimates for $\alpha$ and $\theta$ and an approximate posterior distribution for the latent membership and context variables $\Lambda$ and $Z$. This is accomplished by iterating between E-steps and M-steps as shown in Algorithm \ref{alg:varEM}. The lower bound is given in Equation \ref{eq:ELBO}, but the derivation is left for the appendix.

\begin{equation}\scriptsize \label{eq:ELBO}
\begin{aligned} 
& ELBO =\sum_i^T \log\left[\Gamma\left(\sum_k^K \alpha_k\right)\right]\\
&\quad - \sum_i^T\sum_k^K\log\left[\Gamma(\alpha_k)\right] + \sum_i^T\sum_k^K\left\{(\alpha_k-1)\left[\Psi(\phi_{ik})- \Psi\left(\sum_{k'}^K \phi_{ik'}\right)\right]\right\} \\
&\quad+\sum_i^T\sum_j^J\sum_n^{N_{ij}}\left\{\sum_k^K  \delta_{ijnk}  \left[\Psi(\phi_{ik})- \Psi\left(\sum_{k'}^K \phi_{ik'}\right)\right] \right\} 
\\
&\quad+ \sum_i^T\sum_j^J\sum_n^{N_{ij}}\sum_k^K \delta_{ijkn} \left\{ \left[ \log\left[\theta_{jka(n)_{ij}}\right]\right] - \log\left[1-\sum_{c=0}^{n-1} \theta_{jka(c)_{ij}} \right] \right\}\\
&\quad- \sum_i^T \log\left[\Gamma\left(\sum^K_k \phi_{ik}\right)\right]  + \sum_i^T\sum_k^K\log\left[\Gamma\left(\phi_{ik}\right)\right]
\\
&\quad - \sum_i^T\sum_k^K\left[(\phi_{ik}-1)\left[\Psi(\phi_{ik})- \Psi\left(\sum_{k'}^K \phi_{ik'}\right)\right]\right] \\ &\quad-\sum_i^T\sum_j^J\sum_n^{N_{ij}}\sum_k^K \delta_{ijnk}\log\left[\delta_{ijnk}\right],
\end{aligned}
\end{equation}
$\Gamma(\cdot)$ denotes the gamma function; $\Psi(\cdot)$ denotes the digamma function, the derivative of $\log\Gamma(\cdot)$.

\begin{algorithm}
\caption{Variational EM for Mixed Membership Rank Data \label{alg:varEM}}
\begin{algorithmic}[1]
\State Initialize $\theta^{(0)}$, $\alpha^{(0)}$, K, $\phi^{(0)} = \frac{1}{K}$, $\delta^{(0)} = \frac{1}{K}$ 
\While {(Convergence criterion not yet satisfied)}
	\While {(Convergence criterion not yet satisfied)}
	\Comment{E - Step}
		\State Update $\delta_{ijnk}$
		\State Update $\phi_{ik}$
	\EndWhile
	
	\While {(Convergence criterion not yet satisfied)}
	\Comment{M-Step}
		\State Update $\alpha$ by Newton-Raphson
		\State Update $\theta$ by interior point method
	\EndWhile
\EndWhile
\end{algorithmic}
\end{algorithm}

\subsection{E-Step: Update $\phi$ and $\delta$}
The E-step maximizes the lower bound with respect to the individual level parameters $\phi$ and $\delta$. Taking the derivative of the lower bound yields the following updates for $\phi$ and $\delta$:
\begin{equation} \label{eq:eStepUpdate}\footnotesize
\begin{aligned} \centering
\delta_{ijnk}^{(s+1)} &\propto \exp\left(\Psi(\phi^{(s)}_{ik})- \Psi\left(\sum_{k'}^K \phi^{(s)}_{ik'}\right)  + \log\left[\theta_{jka(n)_{ij}}\right]- \log\left[1-\sum_{c=0}^{n-1} \theta_{jka(c)_{ij}}\right]\right)\\
\phi^{(s+1)} &= \alpha_k + \sum_j^J\sum_n^{N_{ij}}\delta^{(s+1)}_{ijnk}\\
\end{aligned}
\end{equation}
where $\delta$ (a multinomial parameter) is normalized to sum to 1. We continue updating each parameter in a coordinate ascent procedure until the relative increase in the ELBO is below a specified tolerance.

\subsection{M-Step: Update $\alpha$ and $\theta$}
The M-Step, described in Algorithm \ref{alg:thetaUpdate}, fixes the variational parameters $\phi$ and $\delta$, and selects $ \alpha$, the Dirichlet parameter for the membership vectors, and $ \theta$, the Plackett-Luce support parameters, to maximize the lower bound on the marginal log-likelihood. For both parameters, there are no closed form solutions so we use iterative updates. 

For $\alpha$, we use a Newton-Raphson method to maximize the lower bound.
 \begin{equation}\footnotesize
 \begin{aligned}
\frac{\partial ELBO}{\partial \alpha_k} &= T\left(\Psi\left(\sum_{k'}^K \alpha_{k'}\right) - \Psi(\alpha_k)\right) + \sum_i^T \left(\Psi(\phi_{ik}) - \Psi\left(\sum_{k'}^K\phi_{ik'}\right) \right) \\
\frac{\partial ELBO}{\partial \alpha_{k_i} \partial\alpha_{k_j}} &= -T\left( 1_{\{i=j\}}\Psi'(\alpha_{k_i}) - \Psi'\left(\sum_{k'} \alpha_{k'}\right)\right).
\end{aligned}
\end{equation}

Since $\theta$ is subject to the following constraints $\sum_v^{V_j}\theta_{jkv}=1$ and $\theta_{jkv} \geq 0$ for $v=1,\ldots,V_j$, we use an interior point method to select an optimal $\theta$ \citep{wright1999numerical}.
\begin{algorithm} 
\caption{M-step update for $\theta$: interior point method \label{alg:thetaUpdate}}
\begin{algorithmic}[1]
\Require $b_0 > 0$, $M \in \mathbb{Z}^+$, $\theta^{(0)}$
\For{$j \in \left[J\right]$}
	\For{$k \in \left[K\right]$}
		\For {m $\in \left[M\right]$}
			\While {(Convergence criterion not yet satisfied)}
				\State Calculate step direction $\Delta\theta_{jk}$ using penalty term $B_{b_0^m}\left(\theta_{jk} \right)$
				\State Use backtracking line search to determine step length $\tau$
				\State Set $\theta_{jk}^{(t+1)} = \theta_{jk}^{(t)} + \tau \Delta\theta_{jk}$
			\EndWhile
		\EndFor
	\EndFor			
\EndFor
\normalsize
\end{algorithmic}
\end{algorithm}
Because the constraints are only enforced on each individual set $\{\theta_{jkv}\}_{v=1:V_j}$ and the objective function separates into additive terms (with respect to the $\theta_{jk}$), we can select $\theta_{jk}$ for each $\{j,k\}$ separately by solving the minimization problem:
\begin{equation}\label{eq:eqConstr}\small
\begin{aligned}
\min_{\theta_{jk}} & \; \; -ELBO(\theta_{jk}) + B\left(\theta_{jk}\right) &   \;   &\text{subject to } 
\sum_v^{V_j} \theta_{jkv} =1
\end{aligned}
\end{equation}
where $B(\theta)= \infty \sum_v^V 1_{\left[ \theta_{jkv} < 0 \right]}$ (i.e., the non-negativity constraint on $\theta$ has been converted into a penalty term which assigns infinite loss to infeasible points). Because B is not a smooth function of $\theta$, we approximate it with the smooth function $B_b = \frac{-1}{b}\sum_v^{V_j} \log\left[\theta_{jkv}\right]$
and solve the relaxed minimization problem instead. 

For notational ease, we use $\Phi_{jk}(\theta_{jk})$ to denote the objective function; $H$ denotes Hessian of $\Phi_{jk}$; $g$ denotes the gradient of $\Phi_{jk}$ and $\textbf{1}$ denotes a row vector of 1's with length $V_j$.
\begin{equation}\label{eq:derivative} \scriptsize
\begin{aligned}
\frac{\partial \Phi_{jk}(\theta) }{\partial \theta_{jkv_1}} &= -\sum_i^T\sum_n^{N_{ij}} \delta_{ijnk}\left[ \frac{1_{\{X_{ijn} = v_1\}}}{\theta_{jkv_1}} + \frac{\sum_{c=1}^{n-1} 1_{\{X_{ijc} = v_1\}}}{1-\sum_{c=0}^{n-1}\theta_{jva(c)_{ij}}} \right] - \frac{1}{b \theta_{jkv_1}}\\
\frac{\partial^2 \partial \Phi_{jk}(\theta) }{\partial \theta_{jkv_2}\partial \theta_{jkv_1}} 
&= 
-\sum_i^T \sum_n^{N_{ij}} \delta_{ijnk}\left[\frac{\sum_{c=1}^{n-1} 1_{\{X_{ijc} = v_1\}}\sum_{c=1}^{n-1} 1_{\{X_{ijc} = v_2\}}}{\left(1-\sum_{c=0}^{n-1} \theta_{jka(c)_{ij}}\right)^2} \right] \\
\frac{\partial^2 \partial \Phi_{jk}(\theta)}{\partial \theta_{jkv_1}^2}
&=
-\sum_i^T \sum_n^{N_{ij}} \delta_{ijnk}\left[ \frac{-1_{\{X_{ijn} = v_1\}}}{\theta_{jkv_1}^2} + \left(\frac{
\sum_{c=1}^{n-1} 1_{\{X_{ijc} = v_1\}}}{1-\sum_{c=0}^{n-1} \theta_{jka(c)_{ij}}}\right)^2  \right] + \frac{1}{b \theta_{jkv_1}^2}.
\end{aligned}
\end{equation}
Satisfying the Karush-Kuhn-Tucker conditions with the remaining equality constraint yields the update direction $\Delta \theta$ for $\theta^{(s+1)} = \theta^{(s)} + \Delta\theta$ , where
\begin{equation}\label{eq:newtonUpd}\small
\Delta\theta = -H^{-1}\left(g-\textbf{1}^T\frac{\textbf{1}H^{-1}g}{\textbf{1}H^{-1}\textbf{1}^T}\right).
\end{equation}

Because the Newton step in Equation \ref{eq:newtonUpd} uses a quadratic approximation of the objective function, the proposed $\Delta \theta$ increment
may be ill-sized. If the step size is too large, the increment may actually lead to a larger value of the objective function or infeasible updates where $\theta_{jkv}<0$. Thus, we use a backtracking line search, detailed in Algorithm \ref{alg:lineSearch}, to ensure that each update will always increase the lower bound and remain in the feasible set. 
\begin{algorithm}
\caption{M-step update for $\theta$: backtracking line search \label{alg:lineSearch}}
\begin{algorithmic}[1]
\Require $\theta_{jk}$, $\Delta \theta_{jk}$, $\tau_0 \in (0,1)$
\State s = 0
\While {Any($\theta_{jkv} + \tau_0^s \Delta \theta_{j,k} < 0$) or $\left[\text{ELBO}\left(\theta_{jkv} + \tau_0^s \Delta \theta_{jk} \right)< \text{ELBO}\left(\theta_{jkv} \right)\right]$}
	\State s++
\EndWhile
\State \Return $\tau = \tau_0^s$
\end{algorithmic}
\end{algorithm}

\subsection{Algorithm Discussion}
For numerical stability, we first solve the minimization for a small value of $b$, and then use that solution to initialize subsequent minimizations with increasingly larger values of $b$ \citep{wright1999numerical}. In addition, if a particular variable has a large number of options, inverting the $V_j \times V_j$ Hessian can become computationally expensive.  A quasi-Newton or gradient ascent procedure may require less overall computation by avoiding large matrix inversions. We also note that the sub-routines for $\alpha$ and each $\theta_{jk}$ can be computed completely in parallel which may be fruitful if the number of subgroups or variables is large.

As with almost all variational methods, the objective function is multi-modal, so only convergence to a local maximum is guaranteed. Using available prior knowledge can be very helpful in determining reasonable initializations for a specific problem; however, multiple initialization points are recommended to increase the probability of finding the global maximum. 

We found empirically that initializing $\alpha$ and $\theta$ with the following heuristically driven two step procedure generally resulted in stationary points with a larger ELBO. Using a random initialization of $\theta$ and $\alpha$ and setting all values of $\phi$ and $\delta$ to $1/K$, iterate the variational EM procedure until reaching a stationary point $\tilde \alpha$ and $\tilde \theta$. Then, use the resulting global parameters $\tilde \alpha$ and $\tilde \theta$ to initialize a second run (with $\phi$ and $\delta$ reset to $1/K$). The result of the first run is only a stationary point with respect to all the parameters (both global and individual), so resetting the $\phi$ and $\delta$ parameters will generally result in a new stationary point where the global parameters $\hat \theta$ and $\hat \alpha$ are different than the intermediate initialization points.

Because the dimension of even just the $\alpha$ and $\theta$ parameters can be quite large, a huge number of random restarts may be needed to explore the space well when selecting uniformly. We posit that using this two step procedure to find initialization points concentrates the search in areas where the ELBO is likely to be larger.

\section{Eurobarometer Analysis} \label{sec:interpretation}
We now analyze rank data from the Eurobarometer 34.1 survey \citep{Reif2001}. We removed individuals with missing data (i.e. anyone who did not respond to all 3 questions of interest) and individuals who had reported ties in any of their rankings, leaving 11,872 individuals of the original 12,733. 

\subsection{Model Selection} 
Table \ref{tab:topFive} shows the top 5 observed rankings for each question. In particular, we observe that the response which ranks the policy priorities in the exact order of presentation is the most common pattern for drug priorities and is also the $6^{th}$ and $28^{th}$ most common pattern for alcohol and AIDS. If some individuals used this ordering out of convenience, the resulting responses may not be informative of true policy preferences. To capture this tendency, we include a subgroup whose preferences correspond with the presentation-ordered permutation. Following \cite{gormley2006analysis}, we also include a ``noise" subgroup whose preferences are uniform across all policy priorities. We fix the support parameters, $\theta$, for these groups, but estimate their relative frequencies (the corresponding elements of $\alpha$). This approach is similar to the extended grade of membership model \citep{erosheva2007describing} which also models specific response patterns with unusually high counts; however, here we allow for partial membership in the fixed subgroups, whereas \citet{erosheva2007describing} assume that some individuals are full members of the fixed subgroups.

We use a held out ELBO procedure to select an appropriate number of subgroups $K$. We randomly split the sample in half to create a training set and test set. For each $K = 3, 4, \ldots 10$ (we do not include $K = 2$ because that would only be the 2 fixed groups), we first fit a model to the training set and select global parameters $\hat \alpha$ and $\hat \theta$. Then, to compute the held out ELBO on the test set, we use a single E-step which fits the individual variational parameters $\phi$ and $\delta$ for the test set given the $\hat \alpha$ and $\hat \theta$ from the training set. We use 40 different initialization points $\theta_0 \sim \text{Dir}(\{a, a, \ldots a\})$ at each $a = .6, 1.1,\text{ and } 1.5$ and select the stationary point across all $K$ with the highest resulting held out ELBO. We then used the stationary point selected by the procedure to initialize a final run with the results presented below.

\begin{figure}[htbp!]
\centering
\includegraphics[scale = .42]{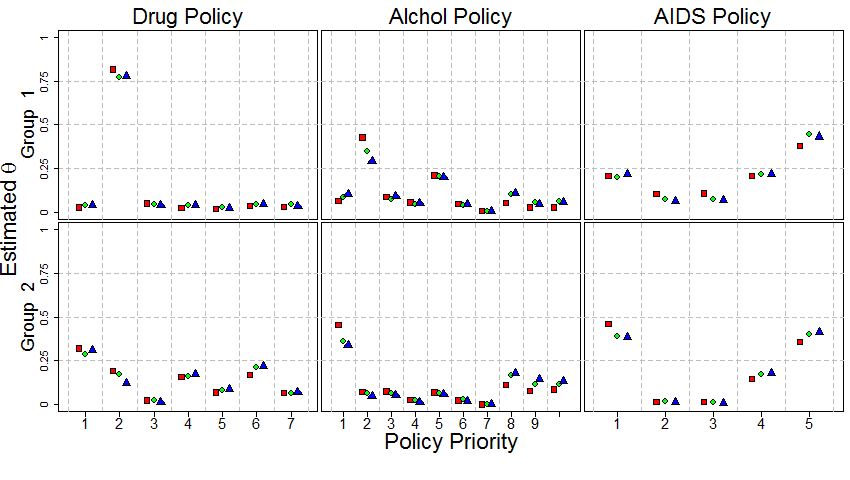}
\caption{\label{fig:batches}The estimated support parameters of two largest subgroups of the selected 5 subgroup model from batch 1 (red squares) is plotted against the estimated support parameters from batch 2 (green circles) and batch 3 (blue triangles).}
\end{figure}

To ensure that the model interpretation is not dependent on the selected training/test set, we repeated this procedure with 3 different training/test splits. For each batch, the held out ELBO values do vary widely within a fixed $K$ due to the multi-modality of the ELBO. However, we see the same trend in all 3 cases; the largest held out ELBO values for each $K = 3,4,5,6$ are somewhat close and peak at either 4 or 5 and the maximum held out ELBO values decrease rapidly as $K$ increases beyond 6. Of the three batches, the first batch selects a 5 subgroup model (including the two fixed groups) and the other two batches select a 4 subgroup model. Figure \ref{fig:batches} shows that the two largest subgroups of the first batch (5 subgroup model) are very similar in structure to the two non-fixed subgroups of the second and third batches (4 subgroup models). Since the 5 subgroup model has the largest ELBO, we describe that model in the remainder of this article. 

\subsection{Goodness of Fit}
\begin{figure}[ht!]
\centering
\includegraphics[scale = .24]{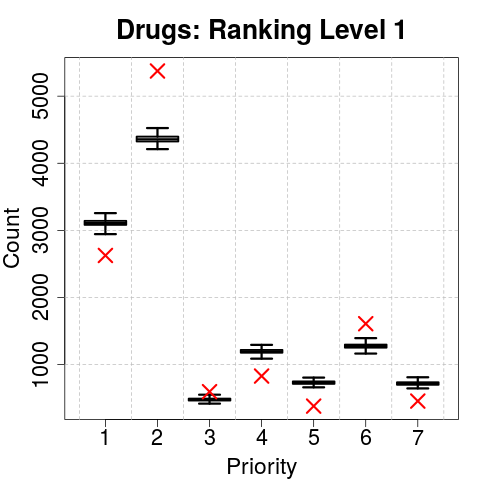}
\includegraphics[scale = .24]{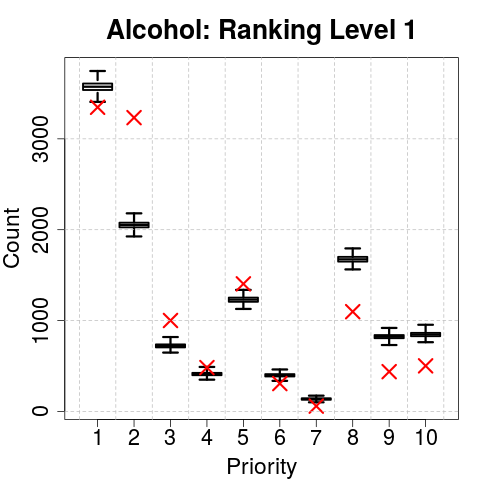}
\includegraphics[scale = .24]{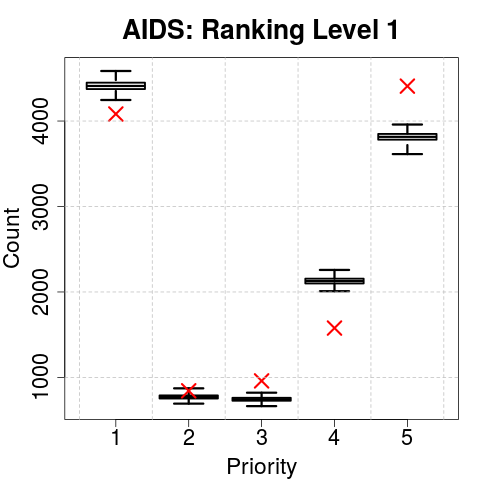}
\caption{\label{fig:goodnessOfFit}1000 simulated top priority rankings using the fitted $\hat \alpha$ and $\hat \theta$ values are shown with the boxplots. The observed counts are indicated by the red x.}
\end{figure}

To check goodness-of-fit, we generate 1000 simulated data sets using the fitted values $\hat \alpha$ and $\hat \theta$. Figure \ref{fig:goodnessOfFit} shows that the model captures the general trend of observed counts for the first place rankings of each variable. These plots are not quite posterior predictive checks because $\alpha$ and $\theta$ are fixed so the variability of the simulated outcomes is smaller than if $\alpha$ and $\theta$ were also considered random quantities. 

\subsection{Model Interpretation}
\begin{figure}[htbp!]
 \centering
\includegraphics[scale = .41]{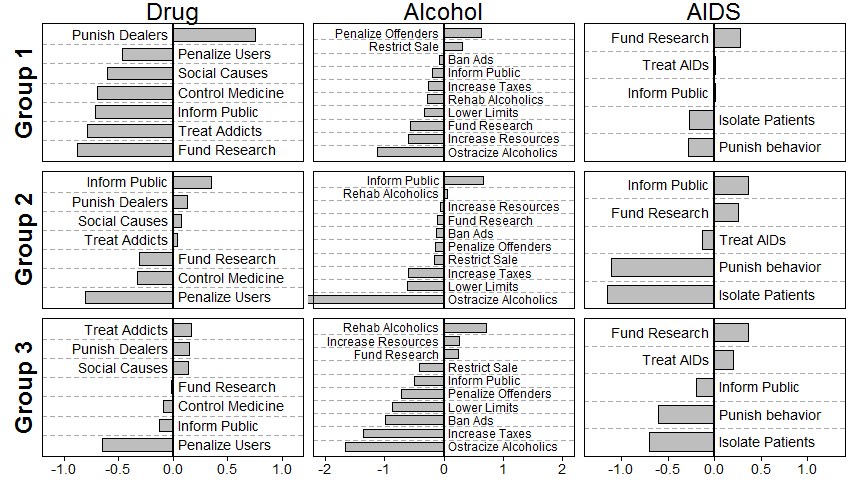}
\caption{\label{fig:tornado}Each barplot in the figure shows $\log_{10}\left(\hat \theta_{jkv}/(1/V_j)\right)$ so that priorities which are more likely to be selected than average have a positive bar height and priorities which are less likely to be selected than average have a negative bar height. The priorities for each subgroup are reordered vertically by largest estimated support to smallest estimated support. Note that the bar for ``Ostracize Alcoholics" for subgroup 2 has been truncated.}
\end{figure}

Table \ref{tab:supParam} presents the ratio of the estimates $\hat \theta$ and uniform support parameters (ie, $\hat \theta_{jkv} / \frac{1}{V_j}$). Thus, the reported values represent how many times more likely a full member of a subgroup would be to select a specific policy as their top priority compared to an individual selecting policies randomly. A value larger than 1 indicates that the policy is more popular than average for the variable and subgroup, and a value less than 1 indicates that the policy is less popular than average. The $\log_{10}$ of these values are also represented in Figure \ref{fig:tornado} where priorities favored more than average have a positive bar height and priorities favored less than average have a negative bar height. Furthermore, within each subgroup, priorities are sorted by estimated support allowing readers to more easily characterize subgroup preferences.

Subgroup 1 generally favors punitive policies. For illegal drugs, the top two priorities are ``Punish dealers" and ``Penalize users," and subgroup 1 is 44 times more likely to select ``Punish dealers" than the least favored option of ``Funding research." Similarly, for alcohol, the most popular policies are ``Stricter penalties for offenses" and ``Restricting sale." Although ``Ostracizing alcoholics" is the least popular policy for all subgroups, in subgroup 1, it is only 55 times less likely than the top priority while it is 260 times less popular than the top option for subgroup 3 and numerically zero for subgroup 2\footnote{For alcohol, there are 10 options and at most 5 ranking levels. Thus, it may be possible for an option to appear extremely infrequently or not at all.}. For AIDS, although the two punitive options (``Punish behavior" and ``Isolate patients") are the least favored policies, subgroup 1 is only roughly 3.5 times less likely to select these two options relative to the most popular option of ``Funding research." 

Subgroup 2 generally prioritizes ``Information campaigns." ``Information campaigns" are roughly 1.7 times more likely to be selected as the top priority than the second most popular option of ``Penalizing dealers" and 14 times more likely than the least popular option ``Penalize users." For alcohol, ``Information campaigns" are 4 times more likely than the second most popular alternative ``Rehabilitate alcoholics". The least popular option is ``Penalize users", with an estimated support parameter that is numerically zero; the two other least popular policies include ``Increasing taxes" and ``Lowering limits." The dislike for these options is consistent with the idea of limited government social intervention. For AIDS, subgroup 2 is the only subgroup for which ``Funding research" is not the most popular option, with ``Information campaigns" roughly 1.3 times more likely than ``Funding research."  

Finally, subgroup 3 typically supports rehabilitation and treatment, as well as funding research. For illegal drugs, although the most popular policy is ``Treating addicts," ``Punishing dealers" and ``Addressing social causes" are also popular policies. For alcoholism, ``Rehabilitation" is by far the most popular policy with ``Funding research" and ``Increasing resources for rehabilitation" as the only other options with substantial support. For AIDS, this subgroup expresses strong support for ``Treating AIDS" and ``Funding research."

Broadly speaking, the identified groups are consistent with the typical Left (liberal) vs Right (conservative) political ideology archetypes. The focus on punitive measures is consistent with a right leaning approach towards governance while the focus on information and rehabilitation typifies a more left leaning approach \citep{cavadino2006penal}.

\begin{table}[ht] 
\centering
\caption{\label{tab:supParam} The estimated support parameters divided by the support parameter under uniform selection. The bootstrapped 95\% confidence intervals are shown in parentheses.} {\tiny
\begin{tabular}{|l|ccc|}
  \hline
 & Subgroup 1 & Subgroup 2 & Subgroup 3 \\
 & \textit{Punitive Subgroup} & \textit{Info Subgroup} & \textit{Rehab Subgroup}\\ 
  \hline 
  & & & \\[-.5em]
   &\multicolumn{3}{c|}{\textbf{Drug Policy}} \\       & & & \\ [-.5em]
Inform Public & 0.191 (0.175, 0.209) & 2.249 (2.17, 2.324) & 0.745 (0.694, 0.8) \\ 
  Punish Dealers & 5.719 (5.623, 5.808) & 1.338 (1.27, 1.404) & 1.411 (1.304, 1.527) \\ 
  Penalize Users & 0.344 (0.316, 0.376) & 0.157 (0.147, 0.167) & 0.225 (0.204, 0.246) \\ 
  Treat Addicts & 0.162 (0.149, 0.176) & 1.103 (1.064, 1.144) & 1.456 (1.373, 1.542) \\ 
  Fund Research & 0.132 (0.122, 0.143) & 0.489 (0.469, 0.509) & 0.966 (0.913, 1.017) \\ 
  Social Causes & 0.251 (0.231, 0.272) & 1.197 (1.149, 1.247) & 1.386 (1.313, 1.461) \\ 
  Control Medicine & 0.202 (0.186, 0.219) & 0.468 (0.449, 0.489) & 0.81 (0.767, 0.853) \\ 
    \hline 
      & & & \\ [-.5em]
     &\multicolumn{3}{c|}{\textbf{Alcohol Policy}} \\       & & & \\ [-.5em]
Inform Public & 0.648 (0.609, 0.689) & 4.565 (4.421, 4.709) & 0.324 (0.296, 0.355) \\ 
  Penalize Off & 4.269 (4.104, 4.432) & 0.722 (0.681, 0.767) & 0.192 (0.168, 0.218) \\ 
  Ban Ads & 0.85 (0.798, 0.904) & 0.744 (0.702, 0.787) & 0.104 (0.089, 0.121) \\ 
  Inc Taxes & 0.541 (0.509, 0.576) & 0.25 (0.23, 0.271) & 0.045 (0.036, 0.055) \\ 
  Rest Sale & 2.086 (1.994, 2.181) & 0.7 (0.662, 0.741) & 0.391 (0.347, 0.438) \\ 
  Low Limits & 0.472 (0.445, 0.5) & 0.243 (0.224, 0.262) & 0.134 (0.117, 0.154) \\ 
  Ostze Alcs & 0.074 (0.065, 0.084) & 0.00 (0.00, 0.00) & 0.022 (0.017, 0.028) \\ 
  Rehab Alcs & 0.535 (0.505, 0.563) & 1.128 (1.085, 1.171) & 5.194 (5.011, 5.38) \\ 
  Fund Research & 0.27 (0.25, 0.288) & 0.787 (0.751, 0.822) & 1.788 (1.69, 1.887) \\ 
  Inc Resources & 0.254 (0.234, 0.274) & 0.862 (0.825, 0.898) & 1.805 (1.718, 1.893) \\ 
    \hline   & & & \\ [-.5em]
     &\multicolumn{3}{c|}{\textbf{AIDS Policy}} \\       & & & \\ [-.5em]
Inform Public & 1.026 (0.984, 1.067) & 2.314 (2.234, 2.393) & 0.646 (0.598, 0.7) \\ 
  Punish behavior & 0.521 (0.495, 0.548) & 0.079 (0.072, 0.086) & 0.251 (0.228, 0.276) \\ 
  Isolate Patients & 0.537 (0.508, 0.57) & 0.072 (0.065, 0.079) & 0.199 (0.176, 0.225) \\ 
  Treat AIDS & 1.031 (0.993, 1.068) & 0.74 (0.705, 0.775) & 1.607 (1.524, 1.685) \\ 
  Fund Research & 1.885 (1.819, 1.951) & 1.796 (1.725, 1.869) & 2.296 (2.195, 2.4) \\ 
   \hline
\end{tabular}}
\end{table}

\begin{figure}[ht]
\includegraphics[scale = .3]{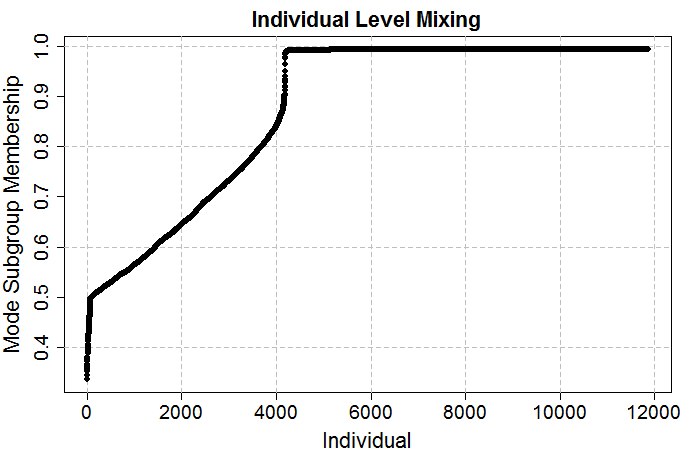}
\caption{\label{fig:modeMem} The estimated degree of membership in each individual's modal subgroup.}
\end{figure}

Shown in Table \ref{tab:alpha}, the small magnitude of $\hat \alpha$, the Dirichlet membership parameter, suggests relatively low levels of intra-individual mixing. However, the modal grade of membership in Figure \ref{fig:modeMem} shows that a quarter of all individuals still exhibit significant intra-individual mixing. We also see that the non-trivial relative frequency estimates of the non-informative fixed groups justify their inclusion in the analysis. 

\begin{table}[ht]
\centering
\tiny
\caption{\label{tab:alpha} The estimated $\alpha$ Dirichlet membership parameter and the 95\% bootstrapped confidence intervals. The estimated relative frequency of each subgroup is given by $\frac{\alpha_{k}}{\sum_{k'} \alpha_k'}$.}
\begin{tabular}{|l|c|c|}
  \hline
  & $\hat \alpha$ & Relative Frequency \\ \hline
Subgroup 1 & 0.05 (0.048, 0.053) & 0.322 (0.314, 0.329) \\ 
  Subgroup 2 & 0.048 (0.047, 0.051) & 0.31 (0.302, 0.318) \\ 
  Subgroup 3 & 0.024 (0.023, 0.026) & 0.154 (0.149, 0.16) \\ 
  Subgroup 4 & 0.014 (0.013, 0.015) & 0.088 (0.084, 0.092) \\ 
  Subgroup 5 & 0.02 (0.019, 0.021) & 0.126 (0.122, 0.131) \\ 
   \hline
\end{tabular}
\end{table}

The Dirichlet distribution for $\lambda_i$ implicitly enforces negative dependence between subgroup memberships, although positive dependencies could be modeled using distributions considered by \citet{lafferty2005correlated}. However, the magnitude of correlations between subgroup memberships is still informative. In Table \ref{tab:membershipCorrel}, the estimated membership in the informative subgroups is more strongly correlated with membership in the other informative subgroups (the least negative correlation between subgroups 1-3 is -.29) than membership in the fixed groups (the most negative correlation between subgroups 1-3 vs subgroups 4-5 is -.25). This is not surprising because subgroups 1-3 indicate a particular ideology on public policy, while subgroups 4 and 5 essentially represent the lack of preferences which align with the dominant subgroups.

\begin{table}[ht]
\centering 
\caption{\label{tab:membershipCorrel}Correlation between the estimated subgroup memberships for each subgroup.} \tiny
\begin{tabular}{c|ccc|cc}
  \hline
 & Subgroup 1 & Subgroup 2 & Subgroup 3 & Subgroup 4 & Subgroup 5 \\ 
  \hline
Subgroup 1 & 1.00 & -0.55 & -0.29 & -0.25 & -0.07 \\ 
  Subgroup 2 & -0.55 & 1.00 & -0.36 & -0.25 & -0.18 \\ 
  Subgroup 3 & -0.29 & -0.36 & 1.00 & -0.16 & -0.03 \\ \hline
  Subgroup 4 & -0.25 & -0.25 & -0.16 & 1.00 & -0.04 \\ 
  Subgroup 5 & -0.07 & -0.18 & -0.03 & -0.04 & 1.00 \\ 
   \hline
\end{tabular}
\end{table}

\subsection{Uncertainty Estimates}\label{sec:bootstrap}
Because $\hat \alpha$ and $\hat \theta$ were selected through a pseudo-MLE procedure, there are no readily available model-based standard errors; however, we estimate standard errors via an empirical bootstrap procedure. For each bootstrap sample $b = 1, \ldots 5000$, we select 11,872 individuals with replacement from the observed sample and use the variational EM procedure to select pseudo-MLE estimates $\hat \alpha^{(b)}$ and $\hat \theta^{(b)}$. Each bootstrap sample run is initialized at the same starting points used for the full model. This initialization avoids overestimating variability in stationary points due to multi-modality of the objective function and seeks to only capture sampling variability. To form 95\% confidence intervals, we take the .025 and .975 quantiles of the bootstrapped estimates.

\subsection{Multivariate vs Univariate Model}
We also acknowledge the implicit decision to use a multivariate model instead of fitting a univariate model for each question. Under a univariate models, subgroup membership $\lambda$ for each individual is estimated independently of the responses to other questions. By contrast, in a multivariate specification, individuals can still exhibit a different mix of subgroups across each question and ranking level, but the posterior estimates are shrunk towards the individual's overall membership $\lambda_i$. 

As a sensitivity analysis, we fit univariate models for each question.  Fewer subgroups may be necessary when considering a univariate model when compared a multivariate model, but since the univariate models are only used to validate the structure of subgroups identified with multivariate data, we fit models with 5 subgroups and do not repeat the model selection procedure. The estimated support parameters $\theta$ do not differ substantially from the multivariate model, but the estimated membership parameters $\alpha$ differ across univariate models in an informative way. Table \ref{tab:alphaEstimates} shows a much higher proportion of membership in the information subgroup for the AIDS univariate model than we see in the drugs univariate, alcohol univariate, or full multivariate models. This suggests that, on average, individuals have a stronger preference for ``Information campaigns" to address AIDS, which is not seen as strongly in addressing alcohol or drugs. As expected, the relative frequencies when averaged across all three univariate models are similar to the relative frequencies of the full  model.

\begin{table}[ht]
\centering \tiny
\caption{\label{tab:alphaEstimates} The estimated $\hat \alpha$ and corresponding relative frequency for each of the univariate and full multivariate models.}
\begin{tabular}{c|ccccc}
  \hline
 & Subgroup 1 & Subgroup 2 & Subgroup 3 & Subgroup 4 & Subgroup 5 \\ 
  \hline
Drug Univariate Est & 0.031 & 0.020 & 0.015 & 0.010 & 0.011 \\ 
  Alcohol Univariate Est & 0.062 & 0.042 & 0.042 & 0.021 & 0.027 \\ 
  AIDS Univariate Est & 0.006 & 0.060 & 0.018 & 0.014 & 0.012 \\ 
  Full Model Est & 0.050 & 0.048 & 0.024 & 0.014 & 0.020 \\ \hline
  Drug Univariate Rel Freq & 0.362 & 0.229 & 0.169 & 0.113 & 0.128 \\ 
  Alcohol Univariate Rel Freq & 0.321 & 0.217 & 0.218 & 0.107 & 0.138 \\ 
  AIDS Univariate Rel Freq & 0.052 & 0.549 & 0.160 & 0.126 & 0.113 \\ 
  Full Model Rel Freq & 0.322 & 0.310 & 0.154 & 0.088 & 0.126 \\ 
  Average Univariate Relative Frequency & 0.259 & 0.319 & 0.181 & 0.115 & 0.127 \\ 
   \hline
\end{tabular}
\end{table}

\subsection{Individual Membership Estimates}
We examine two specific individuals that illustrate the richness of description afforded by using a mixed membership model.

Table \ref{tab:response1115} shows the observed responses from a 68 year old British male. For addressing drugs, his responses follow the presentation ranking, but for alcohol and AIDS, he indicates a preference for research and rehabilitation. We estimate the membership for this man to be 66\% subgroup 3 (Rehab and Research) and 34\% subgroup 5 (Presentation Ordering). Because the mixed membership framework allows for intra-individual mixing, the rank ordered response for drug policy is attributed to the non-informative subgroup. This contrasts with a finite mixture model approach, which would otherwise include this noisy response for drug policy in the estimates for subgroup 3.
\begin{table}[ht]
\tiny
\caption{\label{tab:response1115} Observed Responses from a 68 year old British male.}
\begin{tabular}{rlll}
  \hline
 & Drug Policy & Alcohol Policy & AIDS Policy \\ 
  \hline
Priority 1 & Inform Public & Fund Research & Isolate Patients \\ 
  Priority 2 & Punish Dealers & Increase Resources & Treat AIDS \\ 
  Priority 3 & Penalize Users & Rehabilitate Alcoholics & Fund Research \\ 
  Priority 4 & Treat Addicts & Ban Advertisements & Inform Public \\ 
  Priority 5 & Fund Research & Inform Public & Punish behavior \\ 
  Priority 6 & Fight Social Causes &  &  \\ 
  Priority 7 & Control Medicine &  &  \\ 
   \hline
\end{tabular}
\end{table} 

Table \ref{tab:response6} shows the responses of a 40 year old Spanish female. Her perspectives on alcohol policy differ drastically from her preferences for drugs and AIDS. We see that her top policies for drugs and AIDS are highly punitive, but information campaigns and rehabilitation are preferred for alcoholism. These different perspectives are captured in the model with an estimated membership of 63\% in subgroup 1 and 37\% in subgroup 2.

\begin{table}[ht]
\tiny
\caption{\label{tab:response6} Observed responses from a 40 year old Spanish female.}
\begin{tabular}{rlll}
  \hline
 & Drug Policy & Alcohol Policy & AIDS Policy \\ 
  \hline
Priority 1 & Punish Dealers & Inform Public & Punish behavior \\ 
  Priority 2 & Penalize Users & Rehabilitate Alcoholics & Isolate Patients \\ 
  Priority 3 & Treat Addicts & Restrict Sale & Treat AIDS \\ 
  Priority 4 & Inform Public & Ban Advertisements & Fund Research \\ 
  Priority 5 & Fight Social Causes & Increase Resources & Inform Public \\ 
  Priority 6 & Control Medicine &  &  \\ 
  Priority 7 & Fund Research &  &  \\ 
   \hline
\end{tabular}
\end{table} 

\subsection{Membership by Demographic Subgroup}
The broad interpretation of our results agree with previous studies which have identified demographic characteristics associated with general dispositions toward penal ideology. To examine these demographic trends clearly, we filter out individuals whose membership in subgroups 4 and 5 (the non-informative subgroups) is over 50\%. This leaves 10,448 of the  11,872 original individuals. We then examine the conditional membership of the remaining individuals in subgroups 1,2 and 3 (ie, $\tilde \lambda_i = \frac{\lambda_i}{\lambda_1 + \lambda_2 + \lambda_3}$).

We first examine a self reported measure of Left vs Right political ideology. Individuals were asked: ``In political matters, people talk of the left or the right. How would you place your views on this scale?" In the recorded scale, 1 indicates far left and 10 indicates far right. This is not a perfect analog since each individual likely responded in reference to their national definition of ``center" whereas the subgroup membership estimated from the rank data is a global measure. Nonetheless, we see that there is a very significant Spearman's rank-order correlation of .15 (p-value < 2e-16) between self-reported Left vs Right score from Eurobarometer 34.1 and membership in the ``Punitive subgroup."  

\begin{table}[ht]
\centering
\tiny
\caption{\label{tab:religion} Average conditional membership by denomination and regularity of attending religious services. Regular attendance indicates attending a religious service at least once a week; irregular attendance indicates attending a religious service less than once a week.}
\begin{tabular}{llrrr}
  \hline
  Religion & Attendance & Subgroup 1 & Subgroup 2 & Subgroup 3 \\ 
  \hline
None & NA & 0.36 & 0.45 & 0.19 \\ \hline
  Orthodox & Irregular & 0.21 & 0.58 & 0.21 \\ 
  Orthodox & Regular & 0.30 & 0.47 & 0.23 \\ \hline 
  Protestant & Irregular & 0.40 & 0.40 & 0.20 \\ 
  Protestant &Regular & 0.53 & 0.31 & 0.16 \\ \hline
  Roman Catholic & Irregular & 0.40 & 0.39 & 0.21 \\ 
  Roman Catholic & Regular & 0.47 & 0.34 & 0.18 \\ 
   \hline
\end{tabular}
\end{table}

We also examine the average membership across religious affiliation. Some speculate that Anglo-Saxon cultures are particularly punitive because of Protestant religions with strong Calvinistic overtones \citep{tonry2007determinants} or fundamentalist beliefs \citep{grasmick1992protestant}.
As shown in Table \ref{tab:Demo}, almost all individuals in the survey report their religion as either Roman Catholic, None, Protestant or Orthodox. In addition to denomination, the Eurobarometer also recorded how often an individual attends religious services. We collapse the original categories of ``Several times a week" and ``Once a week" to a single ``Regular attendance" category and collapse ``Few times a year," ``Once a Year," and ``Never" into an ``Irregular attendance" category. Table \ref{tab:religion} shows that those who attend religious services regularly have a much higher average membership in subgroup 1. Also, Roman Catholics and Protestants are much more likely to belong to subgroup 1 than individuals who report no religion or Orthodox Christians. We note that roughly 90\% of the individuals who responded as Orthodox Christians were Greek and roughly 97\% of Greek respondents reported their religion as Orthodox Christianity. This confounding may be the cause of the particularly low subgroup 1 membership for Orthodox Christians. 

In addition, we examine the average estimated membership across levels of education. The Eurobarometer asks ``How old were you when you finished full-time education?" 
%
The average membership in subgroup 1 (punitive) decreases steadily as education increases, a finding that is consistent with previous research on industrialized countries, including Western Europe \citep[e.g.][]{mayhew2002cross,kitschelt2014occupations}.
\begin{table}[ht]
\centering 
\tiny
\caption{Average conditional membership by ``Last age of formal education."}
\begin{tabular}{cccc}
 \hline
Last Age of Formal Education & Subgroup 1 & Subgroup 2 & Subgroup 3 \\ 
  \hline
16 or Less & 0.47 & 0.33 & 0.19 \\ 
  17 to 19 & 0.40 & 0.40 & 0.20 \\ 
  20 to 21 & 0.33 & 0.48 & 0.20 \\ 
  22 or older & 0.26 & 0.54 & 0.20 \\ 
   \hline
\end{tabular}
\end{table}


At the national level, the average memberships are also consistent with qualitative characterizations of national policy. The United Kingdom and Ireland have high average memberships in subgroup 1 (punitive), while Denmark and France are among a cluster of countries with low average memberships in subgroup 1. These findings are generally consistent with previous research using the International Crime Victimization Surveys, which finds punitive attitudes in the United Kingdom and Ireland, and non-punitive attitudes in Denmark and France \citep[e.g.][]{roberts2013public}. In slight contrast to that work, we find Belgium to have a relatively high membership in subgroup 1, and the Netherlands to have a relatively low membership. 

\begin{figure}[ht]
\includegraphics[scale = .4]{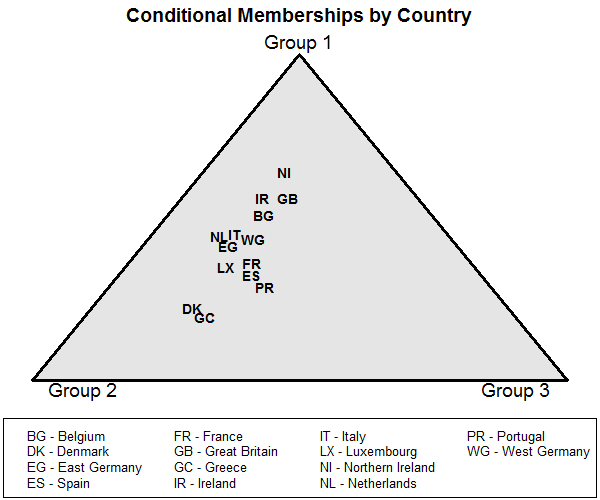}
\caption{\label{fig:countySimplex} The estimated average membership of each country where the non-informative fixed groups (subgroups 4 and 5) have been marginalized out.}
\end{figure}

\section{Discussion}\label{sec:discussion}
In this article, we propose a mixed membership model for multivariate rank data and develop a variational EM estimation approach that is a computationally attractive alternative to fully Bayesian estimation for large scale rank data. Mixed membership models provide valuable insights into latent structure within a heterogeneous population and allow for a richer description when compared to previous mixture model approaches. When MCMC is tractable for smaller data sets as in \cite{gormley2009grade}, the results provide direct samples from the posterior. Nevertheless, the demands placed on human and computer time to conduct such an analysis can be substantial, and scalability of MCMC methods is poor. Ultimately, a mixed membership analysis of larger data sets necessitate other approaches. Of course, what actually qualifies as ``large scale" or ``big data" is dependent on the complexity of analysis. In rank data, the complexity quickly grows as the number of variables and alternatives increase.


In addition to the computational gains, the proposed method extends the method of \cite{gormley2009grade} to explicitly fit the Dirichlet membership parameter $\alpha$. Unless there is strong prior knowledge about subgroup sizes, this extension can result in better fitting models by directly capturing from the data differences in the subgroup structure and the level of intra-group mixing. A direct comparison of both goodness of fit and computational effort is provided in the supplement \citep{wang2015variational:supp}.

To accommodate multivariate ranked data, our model makes the simplest assumption that all context indicators $Z_{ijn}$ are drawn from the same multinomial distribution governed by a single membership vector $\lambda_i$. An alternative and more complex model might include an additional layer of hierarchy between $\lambda_i$ and $Z_{ijn}$ for each variable $j$. This would allow context indicators $Z_{ijn}$ from separate variables to be drawn from different distributions, while still respecting the multivariate structure. 

There are drawbacks, however, to the variational approximation. Because of the multi-modal objective function, many random restarts should be used. Prior knowledge can be used to select good initialization points, but finding a global maximum is not guaranteed. We propose a two-step procedure for initialization, but addressing multi-modality through stochastic optimization methods \citep{bottou2010large} or placing strong priors on the support parameters to induce ``smoothness" in the ELBO are two natural extensions. 

Also, unlike a full Bayesian specification, the variational EM method does not provide a posterior for the global parameters. Frequentist uncertainty estimates, however, can still be achieved through a bootstrap procedure, but each bootstrapped model must be carefully initialized to avoid overestimating variability. To our knowledge, bootstrapping with variational estimation has not been previously used in the existing literature. 

As with any mixture or mixed membership model, selecting the number of subgroups is difficult. Our model selection procedure involves cross-validation of the held-out ELBO. This procedure, however, can be complicated by the multi-modality of the objective function and the selected model might depend on the specific test and training sets. BIC procedures are also widely used although the theoretical justification does not hold in mixed membership models \citep{airoldi2014handbook}. Alternative approaches include stability-based measures \citep{NIPS2002_2139}, direct goodness-of-fit measures \citep{cohen1983assessing}, and non-parametric model extensions such as those based on Dirichlet processes \citep{teh2006hierarchical}. 

Analyzing the Eurobarometer 34.1 data, we find three informative policy preference subgroups as well as substantial support for a uniform ranking group and a presentation-ordered group. The three informative subgroups primarily favor punitive policies, information campaigns, and rehabilitation and research, respectively. When comparing subgroup membership to educational, religious and national demographic information, we see trends which generally agree with the existing literature. In particular, fewer years of formal education and more religious participation is generally associated with more punitive attitudes towards social issues. In addition, at the national level, average subgroup membership roughly agrees with previous characterizations of national punitive attitudes.

Finally, our analysis has implications for survey development. Because of a sizable presentation-ordered subgroup in our analysis, we recommend randomizing the presentation of choices when collecting rank data to decrease bias due to non-informative responses where respondents rank choices by simply following the presentation order. We also note that the variable with the largest proportion of presentation-ordered responses is the question regarding illegal drugs which also happens to allow up to 7 rankings, while the other two questions only allow up to 5 ranking levels. This observation naturally leads to speculation of whether decreasing the number of ranking levels and cognitive load may ultimately lead to more ``informative" responses.   

Although our analysis focused on issues within political science, sociology, and public health, multivariate rank data can elicit and capture a rich representation of individual preferences. We believe that the proposed methodology will be of broad interest. Psychologists, economists, other social scientists, and marketing professionals who analyze large scale rank data can rely on the proposed methodology to represent large scale ranked preferences with realistic models which are still parsimonious and easily interpretable.

\newpage
\bibliographystyle{imsart-nameyear}
\bibliography{rankData_v6}

\newpage
\appendix
\section*{Appendix}
\small
\subsection*{Derivation of Lower Bound on Marginal Log-Likelihood (ELBO)}
The derivation of the lower bound from equation \ref{eq:Jensen2} is shown here.
The lower bound is
\[\log\left[P(X|\alpha, \theta)\right] \geq \E_Q \left( \log(p(X, Z, \Lambda)) \right) - \E_Q\left(\log(Q(Z, \Lambda))\right) \]

Note that $\Gamma(\cdot)$ denotes the gamma function, $\Psi(\cdot)$ denotes the digamma function, and $f_j$ denotes the Plackett-Luce mass function of variable j. $X_{ij}$ denotes the observation of $N_{ij}$ level rankings and $a(n)_{ij}$ indicates the alternative selected by individual i for variable j at ranking level n. Note that for all multinomial mass functions shown below, the size = 1.

We consider each piece of the lower bound separately. The log likelihood for the complete data is
\begin{equation}
\begin{aligned} \label{eq: logLike}
\log\left[p(X, Z, \Lambda|\alpha, \theta)\right] &= \log\left[\prod_i^T \left( \text{Dir}(\lambda_i|\alpha)\prod_j^J\left[\prod_n^{N_{ij}} \text{mult}(Z_{ijn}|\lambda_i)\right]f_j\left(X_{ij}|\theta, Z_{ij \cdot}\right) \right) \right] \\
&= \sum_{i}^T \log\left[\text{Dir}(\lambda_i|\alpha)\right] + \sum_i^T \sum_j^J\sum_n^{N_{ij}}\log\left[\text{mult}\left(Z_{ijn}|\lambda_i\right)\right] \\
&\quad + \sum_i^T \sum_j^J\log\left[f_j\left(X_{ij}|\theta, Z_{ij\cdot}\right)\right] \\
\end{aligned}
\end{equation}

The expectation of the first term with respect to the variational distribution Q becomes
\begin{equation}
\begin{aligned}
\sum_{i}^T \E_Q \log \left[\text{Dir}(\lambda_i|\alpha)\right] &= \sum_i^T\E_q\left\{ \log \left[\frac{\Gamma[\sum_k^K\alpha_k]}{\prod_k^K \Gamma(\alpha_k)}\prod_k^K \lambda_{ik}^{\alpha_k -1} \right]\right\}\\
 &= \sum_i^T \log\left[\Gamma\left(\sum_k^K \alpha_k\right)\right] - \sum_i^T\sum_k^K\log\left[\Gamma(\alpha_k)\right]\\
 &\quad + \sum_i^T\sum_k^K(\alpha_k-1)\E_q\left\{\log\left[\lambda_{ik}\right]\right\} \\
 &= \sum_i^T \log\left[\Gamma\left(\sum_k^K \alpha_k\right)\right] - \sum_i^T\sum_k^K\log\left[\Gamma\left(\alpha_k\right)\right] \\
 &\quad+ \sum_i^T\sum_k^K\left(\alpha_k-1\right)\left[\Psi(\phi_{ik})- \Psi\left(\sum_k^K \phi_{ik}\right)\right] 
 \end{aligned}
\end{equation}

The expectation of the second term with respect to the variational distribution Q becomes
\begin{equation}
\begin{aligned}
\sum_i^T \sum_j^J\sum_n^{N_{ijr}} \E_Q \left\{\log \text{mult}(z_{ijn}|\lambda_i)\right\} &= \sum_i^T \sum_j^J\sum_n^{N_{ij}} \E_Q\left\{ \log\left[\prod_k^K\lambda_{ik}^{Z_{ijnk}}\right]\right\}\\ 
&= \sum_i^T \sum_j^J\sum_n^{N_{ij}} \sum_k^K \E_Q\left\{ Z_{ijnk} \log(\lambda_{ik})\right\}\\ 
&= \sum_i^T \sum_j^J\sum_n^{N_{ij}} \sum_k^K \E_Q\left\{ Z_{ijnk}\right\} \E_Q\left\{\log\left[\lambda_{i,k}\right]\right\} \\ 
&= \sum_i^T \sum_j^J\sum_n^{N_{ij}} \sum_k^K \delta_{ijnk} \left[\Psi\left(\phi_{ik}\right)- \Psi\left(\sum_{k'}^K \phi_{ik'})\right)\right]
\end{aligned} 
\end{equation}

The third term is 
\begin{equation}
\begin{aligned}
\sum_i^T \sum_j^J& \E_Q\left\{\log\left[f\left(X_{ij}|\theta, Z_{ij\cdot}\right)\right]\right\}\\
 &= \sum_i^T \sum_j^J \E_Q \left\{ \log\left[\prod_n^{N_{ij}} \left[ \prod_k^K \left[\frac{ \theta_{jka(n)_{ij}}}{\left(1-\sum_{c=0}^{n-1} \theta_{jka(c)_{ij}}\right)} \right]^{Z_{ijnk}}\right] \right]\right\}\\ 
&=
\sum_i^T \sum_j^J\sum_n^{N_{ij}} \sum_k^K\E_Q\left\{Z_{ijnk} \left( \log\left[\theta_{jka(n)_{ij}}\right] - \log\left[1-\sum_{c=0}^n \theta_{jka(c)_{ij}}\right] \right)\right\} \\
&=\sum_i^T \sum_j^J\sum_n^{N_{ij}}\sum_k^K \delta_{ijnk} \left( \log\left[\theta_{jka(n)_{ij}}\right] - \log\left[1-\sum_{c=0}^n \theta_{jka(c)_{ij}}\right] \right)
\end{aligned}
\end{equation}

Now for the second term of the ELBO-
\begin{equation}
\begin{aligned}
\E_Q\left\{\log\left[Q\left(Z, \Lambda\right)\right]\right\} &= \E_Q\left\{\log\left[ \prod_i^T \left( \text{Dir}\left(\lambda_i|\phi_i\right)\prod_j^J\prod_n^{N_{ij}}\text{mult}\left(Z_{ijn}|\delta_{ijrn}\right)\right)\right]\right\}\\
&= \E_Q\left\{ \sum_i^T \log\left[\text{Dir}(\lambda_i|\phi_i)\right] + \sum_i^T\sum_j^J\sum_n^{N_{ij}}\log\left[\text{mult}\left(Z_{ijn}|\delta_{ijn}\right)\right]\right\}\\
&=\E_Q\left\{\sum_i \log\left[ \frac{\Gamma\left(\sum_k^K \phi_{ik}\right)}{\prod_k^K \Gamma (\phi_k)} \prod_k^K 
\lambda_{ik}^{\phi_{ik}-1}\right] + \sum_i^T\sum_j^J\sum_n^{N_{ij}}\log\left[\prod_k^K \delta_{ijnk}^{Z_{ijnk}}\right] \right\}\\
&= \E_Q \left\{\sum_i^T \log\left[\Gamma\left(\sum_k^K \phi_{ik}\right)\right] - \sum_i^T\sum_k^K \log\left[\Gamma\left(\phi_{ik}\right)\right]\right\}\\
&\quad + \E_Q\left\{\sum_i^T\sum_k^K\left(\phi_{ik}-1\right)\log\left[\lambda_{ik}\right] +\sum_i^T\sum_j^J\sum_n^{N_{ij}} \sum_k^K Z_{ijnk}\log\left[\delta_{ijnk}\right]\right\}\\
&= \sum_i^T \log\left[\Gamma\left(\sum_k^K \phi_{ik}\right)\right] - \sum_i^T\sum_k^K\log\left[\Gamma(\phi_{ik})\right]\\
&\quad + \sum_i^T\sum_k^K\left(\phi_{ik}-1\right)\left[\Psi\left(\phi_{ik}\right)- \Psi\left(\sum_k^K \phi_{ik}\right)\right] + \sum_i^T\sum_j^J\sum_n^{N_{ij}} \sum_k^K \delta_{ijnk}\log\left[\delta_{ijnk}\right]
\end{aligned}
\end{equation}

\end{document}